\documentstyle[aps,prl,epsfig]{revtex}
\begin{document}
\draft
\twocolumn[\hsize\textwidth\columnwidth\hsize\csname %
@twocolumnfalse\endcsname

\preprint{LAUR-xxx}
\title{Finite temperature properties of the 2D Kondo lattice model}
\author{K. Haule$^a$, J. Bon\v ca$^{a,b}$ and P. Prelov\v sek$^{a,b}$}
\address{$^a$ J. Stefan Institute, Ljubljana, $^b$FMF, University of Ljubljana 
Slovenia}
\date{\today}
\maketitle
\begin{abstract}\widetext
Using recently developed Lanczos technique we study finite-temperature
properties  of the 2D Kondo  lattice model at  various fillings of the
conduction band.   At  half  filling the quasiparticle   gap governs
physical  properties  of   the chemical   potential   and  the  charge
susceptibility  at small temperatures.    In the intermediate coupling
regime quasiparticle    gap   scales   approximately  linearly    with
Kondo coupling as $\Delta_{qp}/J\sim 0.3$.  Temperature dependence of the
spin susceptibility reveals the existence of two different temperature
scales.   A spin gap in  the  intermediate regime leads to exponential
drop of the spin  susceptibility at low temperatures.  Unusual scaling
of spin susceptibility is found for temperatures above $T_c\ge 0.6 J$.
Charge susceptibility  at  finite  doping reveals  existence  of heavy
quasiparticles. A new low energy scale is found at finite doping.
\end{abstract} 
] 

\narrowtext

\section{Introduction}

The Kondo lattice model is one of the simplest two-band lattice models
of  correlated electrons.  It is   widely used to model heavy  fermion
materials  where weakly interacting   electrons in wide  bands coexist
with almost localized electrons   in unfilled orbitals of actinide  or
rare-earth elements.  In heavy fermion  materials a remarkable variety
of different   phases can be  found  at low temperatures: paramagnetic
metal with   large    quasiparticle  mass,   anti-ferromagnetic    and
ferromagnetic  phases, unconventional superconductivity, etc.  In most
of  this  cases,    strong electron  correlations    represent the key
ingredient  of the theory  that explains the  rich variety of physical
phenomena.

In this work we investigate the Kondo  lattice model, defined on small
two dimensional square lattices  with periodic boundary conditions. We
use the  finite-temperature Lanczos  method \cite{jaklic}.  The  model
can be written as
\begin{equation}
H = -t\sum_{<{\bf ij}>s} c^\dagger_{\bf {i}s} c_{\bf{j}s} + {\rm H.c.} +
J\sum_{\bf{i}} {\bf S_i}{\bf s_i}
\label{kondo}
\end{equation}
where   ${\bf    s_i}   =  \sum_{s  s^\prime }   c^\dagger_{is^\prime}
{\bf\sigma}_{s^\prime s} c_{is}$ and  summation $<{\bf  ij}>$ runs over
nearest neighbors.  There are two distinct types of degrees of freedom
in this model: free electrons described by $c$ operators and localized
spins described by ${\bf S}$.  In the limit when $J=0$ the two systems
are decoupled which leads  to a large degeneracy of  the states due to
noninteracting  spins.  At finite Kondo coupling  $J\not =  0$ the two
systems interact.  It is believed, that the  interplay between the two
degrees of freedom represents the most important physical mechanism of
the heavy Fermion materials.  Associated with the two distinct systems
are   two competing   interactions  that govern   the  low-temperature
physics.    Finite   Kondo  coupling  leads   to   formation  of Kondo
spin-singlets between the conducting electrons and the localized spins
which screens  the moments  of   localized spins.  Singlet   formation
competes with band propagation   of electrons in the  conduction band.
Coupling  between  kinetic energy  of  band electrons and  local Kondo
coupling  leads to  formation  of  the  Ruderman-Kittel-Kasuya-Yoshida
(RKKY) interaction between localized spins. The Kondo screening on the
one hand and  the RKKY  interaction on   the other are  in many  cases
competing interactions.  Conditions,  under   which one or  the  other
prevails depend  mostly  on the  strength of  the  Kondo coupling, the
electron filling and the dimensionality of the system.

A  number of  theoretical approaches  has  been  applied to the
investigation  of the Kondo lattice  model \cite{tsune}. In the strong
coupling     regime     perturbation    theory   can    be     applied
\cite{sigrist}. Large-$N_f$ expansion can be used in the case of large
localized spin-degeneracy   $N_f$ \cite{read}.   Slave-boson  approach
\cite{read}, Gutzwiller    variational treatments   \cite{coleman} and
recently developed   strong coupling    method \cite{eder}   have been
successful in predicting  the heavy  mass  of the quasiparticles,  the
phase  diagram and the properties of  the spectral  functions.

Numerical calculations have  been  mostly limited to one   dimensional
systems where  various   well developed techniques  are available  and
finite-size effects can  be easily controlled.   Calculations on small
systems have demonstrated   that at half-filling the  one  dimensional
Kondo  lattice  model is  a   spin-liquid  with   a  finite  spin  gap
\cite{tsunegap}.   Density-matrix      numerical-renormalization group
(DMRG)  calculations  \cite{white}, provided accurate determination of
the   spin  and charge gaps   as   a function  of   the Kondo coupling
\cite{yu,shibata}. Recently, a powerful finite-temperature DMRG method
\cite{shibata1,wang} have provided  reliable results for thermodynamic
\cite{shiba2,tsune1,shiba} and dynamic  properties \cite{shiba,shiba1}
of the model at $T>0$.

While there  are many reliable  numerical results of the Kondo lattice
model in  1D, much  less  is known  about the model  in two  or  three
dimensions. Based on theoretical considerations conceptually different
physical  behavior is expected    in higher  dimensions.   Spin-charge
separation     exists in  1D models    as   a  consequence of   strong
correlations.  It is reflected in different  energy scales that govern
the low-energy  behavior of spin and charge  excitations  leading to a
difference  between the spin  and  the  charge velocities.   Luttinger
liquid parameters define power-law  behavior of  correlation functions
in  1D while in higher dimensions  exponential behavior of correlation
functions is expected unless   long-range order exists.  The  lack  of
long-range order in 1D is  responsible for Kondo screening to overcome
the RKKY  interaction  for any finite  $J$.   In two  dimensions there
exists a   critical value of $J_c/t\sim   1.4$ \cite{wang1,shi,assaad}
below  which  RKKY   interaction     prevails and  system       orders
antiferromagnetically  in the case  when  the  conduction band is  half
filled.

The main purpose  of this work  is to explore thermodynamic properties
of   the  2D Kondo  lattice  model.    We focus  our investigations to
intermediate and high temperatures and  try to identify various  energy
and temperature scales that govern the spin and the charge response of
the   system.  Due to  small  system size  we are not  able to explore
extremely  low temperatures since  in  this regime strong  finite-size
effects emerge.   For the same  reason we  limit our  calculations to
intermediate and strong  coupling regime where physics is sufficiently
local  so  that our  results remain  valid   even in the thermodynamic
limit.

\section{Results}

Our   numerical  calculations  are    performed  by recently developed
finite-temperature  Lanczos  method   \cite{jaklic}.  We   investigate
square lattices of $N=8$ and $10$ sites. Most of the results presented
are   for   the   $N=10$   case.    Standard  zero-temperature   exact
diagonalization  results on  small  clusters are generally plagued  by
strong   finite-size  effects.   Performing    calculations at  finite
temperatures and within the grand-canonical ensemble gives us not only
the thermodynamical properties of  the  system, but  most  importantly
diminishes   finite-size effects   for   $T>T^*$  \cite{jaklic}.  This
temperature depends  primarily on  the   number of low  lying  excited
states  in the system.  The $T^*$  can be very   small when the system
possesses either: a) a large number  of low-lying energy states, or b)
if physics is sufficiently local.   Local physics is expected at large
Kondo coupling where the size of the Kondo singlet is  of the order of
a few lattice spacings.   We  present results for half-filled  conduction
band    case $n_c=1$  and   at   finite  doping  $\delta$  defined  by
$n_c=1-\delta$.   Due  to particle-hole   symmetry only  $\delta>0$ is
considered.  In this work  we  restrict calculations to  thermodynamic
quantities as are the chemical   potential $\mu$, spin  susceptibility
$\chi_s=<{S^z_{tot}}^2>/T$,            charge           susceptibility
$\chi_c=-d\delta/d\mu=(\langle n_c^2\rangle  -\langle n_c\rangle^2)/T$
and the  specific heath $c_v=-T \partial^2   F/\partial T^2$.  Despite
small system   size we can   compute  thermodynamic quantities at  any
doping $\delta$ simply  by choosing the appropriate chemical potential
\cite{jaklic}.

\subsection{The chemical potential $\mu$}

\begin{figure}[tb]
\begin{center}
\epsfig{file=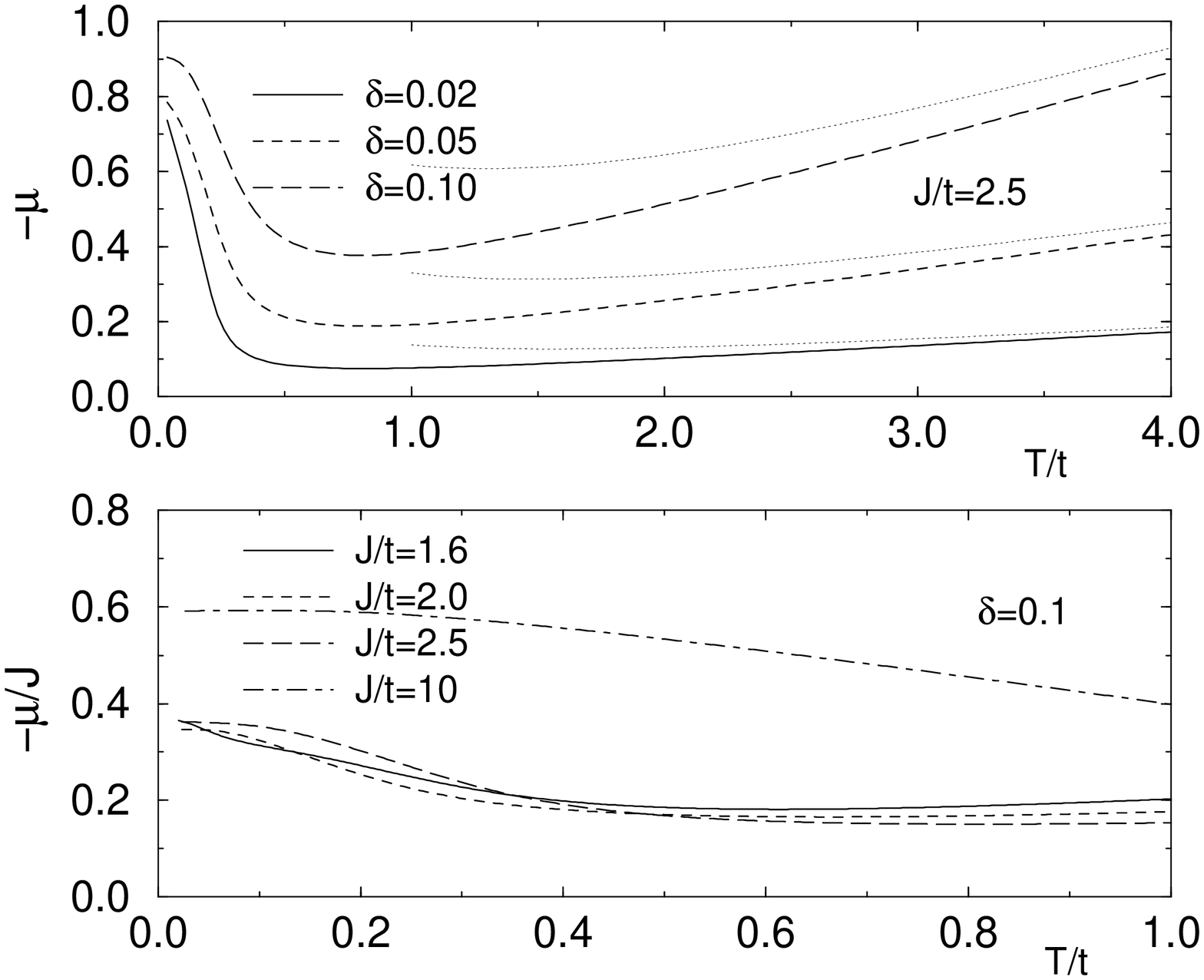,height=80mm,angle=-90}
\end{center}
\caption{ Chemical potential $\mu$ as a function of temperature $T$;
a) at different dopings $\delta$ and fixed $J$, b) at fixed doping
$\delta$ and different couplings $J/t$. }
\label{mu}
\end{figure}

In Fig.~(\ref{mu}a) we show chemical potential $\mu$  as a function of
temperature for small doping values and $J/t=2.5$.  Due to particle-hole
symmetry the relation $\mu(\delta=0)=0$ is  valid at any  temperature.
At   high temperature  and  finite doping   the chemical  potential
approaches a simple expression (faint dotted lines), calculated by the
finite-temperature expansion
\begin{equation}
\mu = T\left ( \log{1-\delta\over 1+\delta} -
{\delta\over 2T^2}\left({3J^2\over 8}+8t^2\right)
\right ).
\label{muft}
\end{equation}
While the first term is model independent,  the second term represents
the  first nontrivial finite-temperature correction. Numerical results
start  deviating significantly  from the  hight-temperature expression
below $T/t\sim 2$. At small temperatures chemical potential approaches
a finite value even in the limit $\delta\to 0$.  Infinitesimally small
doping away  from the half-filled conduction  band  leads to an abrupt
change  of the chemical potential.   Such behavior indicates formation
of the  quasiparticle  gap  $\Delta_{qp}$ also  found in  the  1d DMRG
calculations  \cite{tsune1}.  To investigate the  quasiparticle
gap in more  detail, we  present in  Fig.(\ref{mu}b)  $\mu/J$ at fixed
doping for different  choices of  the  Kondo  coupling strength  on  a
system of $N=10$ sites. We  find that the quasiparticle gap  increases
almost linearly with  the   Kondo  coupling $J$ in  the   intermediate
coupling  range, {\it i.e.} for  $J/t=1.6,2.0$  and 2.5.  Extrapolated
values of the quasiparticle gap in the limit $T\to 0$ are presented in
table (\ref{gaps})   for systems  of  $N=8$  and $10$  sites. Results,
obtained   from   the two systems    agree    reasonably well in   the
intermediate coupling  range. The lack of  finite-size effects  in the
limit   when  $T\to  0$   is   attributed  to   the existence  of  the
quasiparticle gap.  Near   the  strong coupling limit, $J/t=10$,   the
extrapolated values  for  the quasiparticle gap  agree  well with  the
strong coupling  result $\Delta_{qp}={3\over  4}J  -2t +  {13\over  6}
t^2/J=5.72 $\cite{tsune}.

\subsection{Spin and charge susceptibilities}

\subsubsection{Strong coupling (atomic) limit}

\begin{figure}[tb]
\begin{center}
\epsfig{file=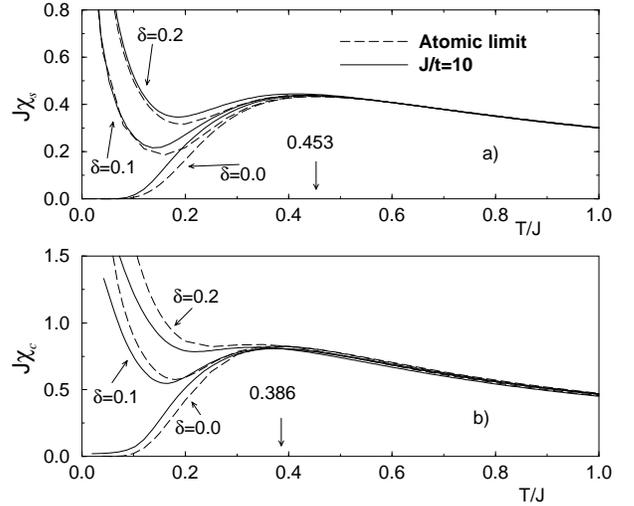,height=80mm,angle=-90}
\end{center}
\caption{ Spin a) and  charge  b) susceptibilities $\chi_s$,  $\chi_c$
vs.   $T/J$ calculated numerically at   $J/t=10$  (full lines). Dashed
curves represent results  obtained  in the atomic limit.  Positions of
the peaks $T_s$ in a) and $T_{qp}$ in b)  in the atomic limit and zero
doping are indicated with arrows.}
\label{strong}
\end{figure}

At large  values  of Kondo coupling,  $J/t=10$,  physics of the  Kondo
model becomes local.   We support   our   claim by results shown    in
Fig.~(\ref{strong})  where we present comparison  of spin (and charge)
susceptibility for $J/t=10$  and analytical  results obtained within
the   atomic limit of   the Kondo  model.    In  this limit the  grand
canonical sum can be calculated as
$Z=\sum_{j=1}^8 e^{-\beta (E_j^{at}-\mu N)},$
where $\beta$ is the  inverse temperature and only  8 states are taken
into account: the singlet state  with the energy $E^{at}_{S=0}=-3J/4$,
the  three-fold  degenerate  triplet   state  $E^{at}_{S=1}=J/4$  both
containing  one  conduction electron and  four-fold  degenerate states
$E^{at}_{S=1/2}=0$ consisted  of   an  empty  and a   doubly  occupied
conduction level each of  them with two different spin configurations.
Values of spin,  quasiparticle and charge  gap are in this limit given
by $\Delta_s/J=1,~\Delta_{qp}/J=3/4,$ and $\Delta_c/J=1.5 $. At zero
doping $(\delta=0)$ spin and charge susceptibility are given by simple
expressions
\begin{eqnarray}
\chi_s &=& \beta {1+2 e^{-\beta J/4}\over 4 + 3 e^{-\beta J/4}
+ e^{3\beta J/4}},\\
\chi_c &=& \beta {4 \over 4 + 3 e^{-\beta J/4} + e^{3\beta J/4}},
\label{chistr}
\end{eqnarray}
where   $\beta$  is     the   inverse    temperature and      $\mu=0$.
Susceptibility, Eq.~(\ref{chistr}) presented in  Fig.~(\ref{strong})
follow $1/T$ law at high temperatures.  At lower temperatures we see a
peak, marked by arrows. We introduce  two temperature scales that mark
the peak  positions:  $T_s/J=0.453$ and $T_{qp}/J=0.386$ for  spin and
charge  susceptibility   respectively.    Given values  were  obtained
analytically.  At low temperatures both susceptibilities approach zero
at $\delta=0$  which is consistent with the  existence of a gap in the
excitation spectrum.  Low-temperature  behavior is in both cases given
by $\chi_{c,s}\propto \beta~{\rm Exp}\left[{-3\beta J/4}\right]$ which
leads  us   to a conclusion,  that   the quasiparticle  (smallest) gap
$\Delta_{qp}/J=3/4$  governs  the   low-temperature  behavior of  both
susceptibilities.  This  is possible since  a quasiparticle excitation
modifies the charge configuration  and  also changes the spin  quantum
number by $\pm 1/2$. Even though  both susceptibilities share a common
gap, they reach a maximum at slightly different temperatures $T_s$ and
$T_{qp}$  which is due to  different nature  of the excitation spectra
above the gap.

We see that numerical  calculations at large $J/t=10$,  also presented
in Fig.~(\ref{strong}),  agree reasonably well with analytical results
at  zero and  finite doping.   Agreement  at finite doping is somewhat
surprising since in the atomic limit  only a single  site of the Kondo
lattice  is taken into account.   To understand the divergence of spin
and charge susceptibilities at low temperatures  and finite doping, we
perform low-temperature expansion which gives us
\begin{eqnarray}
\mu &=& -{3\over 4}J + T\ln\left({2(1-\delta)\over\delta}\right),\\
\chi_s &=& {\delta\over 4 T},\\
\chi_c &=& {\delta(1-\delta)\over T}.
\end{eqnarray}
Divergence  at low  temperatures is in   the strong coupling regime  a
consequence of a degenerate level system.  At even lower temperatures,
numerical results should saturate towards finite values in both cases,
the spin and the charge susceptibility. In the  later case a deviation
from the predicted $1/T$ law can be seen in Fig.~(\ref{strong}b) while
in the   former   case such deviation  is   expected  at  even   lower
temperatures.

\subsubsection{Zero doping}

In Fig.~(\ref{chi}a) we  present spin susceptibilities $J \chi_s(T/J)$
at  zero   doping,  $\delta=0$  for  different   values  of the  Kondo
interaction J.  At  high temperature  $(T>J)$  numerical results agree
with the high-temperature expansion,
\begin{equation}
\chi_s = {3-\delta^2\over 8T}\left[1-{1-\delta^2\over 3-\delta^2}{J\over 2T}
\right],
\label{chis}
\end{equation}
performed  to  the  first  nontrivial   order.  In  the   intermediate
temperature regime we  find   rather  surprising result.  All   curves
merge on a single curve for $T/J>T_c/J\sim  0.6 $.  It could be argued
that this is because  the high-temperature result in Eq.~({\ref{chis})
scales  with  $J$,  {\it     i.e.}    the function $J\chi_s(J/T)$   is
independent of $J$.  However, the agreement with  the Eq.~{\ref{chis})
is only within $10\%$ up to $T/J\sim  1$ (see Fig.~(\ref{chi}a)) while
the overlap of susceptibilities calculated for  a wide range of $J/t$
is within a few percents.
At  low temperature, spin susceptibility reaches  a maximum at $T=T_s$
and then approaches zero. In  the strong coupling   limit spin gap  is
larger than the quasiparticle gap however at  smaller $J/t$ this is no
longer true. In the region of small  $J/t\sim 1.4$ spin gap approaches
zero due to formation   of AFM order \cite{assaad}.   Low  temperature
behavior of spin susceptibility in the intermediate coupling regime is
thus governed by the  spin gap.  There are  two possible approaches to
estimate  the  spin gap  using   our method.  In the  zero-temperature
approach the spin gap equals  the energy difference between the lowest
$(S=0)$  and the first excited  $(S=1)$ state.  At finite temperature,
the spin gap  is roughly proportional to the  position of the peak  in
$\chi_s(T)$ given by the  activation  temperature $T_s$.  We  believe,
that the  second method, even though indirect,  gives results that are
closer to the thermodynamic limit.
\begin{figure}[tb]
\begin{center}
\epsfig{file=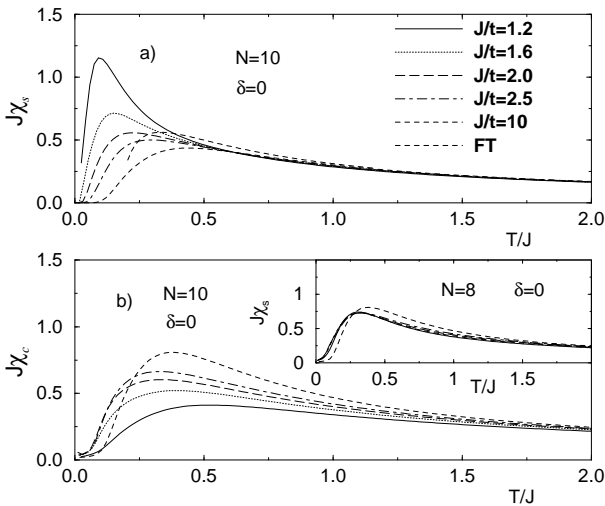,height=80mm,angle=-90}
\end{center}
\caption{ Spin a) and charge  b) susceptibilities $J\chi_s$, $J\chi_c$
vs.  $T/J$  at  zero   doping.  Faint   dashed line in    a) represent
high-temperature  expansion result,  Eq.~(\ref{chis}).  Legends, given
in a) apply also  for b) and the inset  in b). In the  inset in b) all
six different curves for $J/t=1.2\dots 10$ are presented. All, except
for  $J/t=10$  appear as a   single line. Legend  FT   in a) indicates
analytical result in Eq.~(\ref{chis}).}
\label{chi}
\end{figure}

Values of $T_s$  are  presented in Table~(\ref{gaps}).  As   seen from
Fig.~(\ref{chi}a) with increasing coupling $J$, low-temperature peak
in  spin  susceptibility moves toward  higher  values  of $T/J$.  $T_s$
therefore does not scale linearly with  $J$ for small  $J$ as does the
quasiparticle gap (see also Table~(\ref{gaps})). This points towards a
non-linear  dependence  of spin-gap   vs.   $J/t$  that  is  found  in
one-dimensional system \cite{tsune}  and also  recent calculations in
the  2D system \cite{assaad}.  A qualitatively  good agreement is also
found between  $T_s$  and the   spin  gap obtained by  the  projection
quantum Monte Carlo simulations  \cite{assaad}.

At small values of the Kondo coupling $J/t<1.4$ gapless AFM long range
order  develops on  an  infinite  lattice as  a   consequence of  RKKY
interaction \cite{shi,assaad}.    Uniform spin susceptibility  in this
case saturates  around the temperature  which   is given by  the  RKKY
interaction between localized spins. Our results in this regime become
less reliable at low temperature due to strong finite-size effects.
The charge susceptibility $\chi_c$    shows in sharp contrast  to  the
strong  coupling limit in   many respects different  behavior than the
spin susceptibility.    Starting  from the  high-temperature limit  we
first  present    the  high-temperature     result for     the  charge
susceptibility
\begin{equation}
\chi_c = {1-\delta^2\over 2T}\left[1-{1+\delta^2\over 8T^2}
\left({3J^2\over 8}+8t^2\right)
\right].
\label{chic}
\end{equation}
Note  that  in contrast to  spin  susceptibility, the first nontrivial
correction in the inverse  temperature  is $1/T^3$.  The  agreement of
numerical results with the high-temperature expansion strongly depends
on $J$.   In Fig.~(\ref{chi}b) we show $J  \chi_c(T/J)$ at zero doping
and a wide range of $J$. At large  $J/t=10$, where $J$ is the dominant
energy  scale  in   the   system, numerical  results agree   with  the
analytical result, given by Eq.~(\ref{chic}),  down to $T/J=0.4 $.  At
smaller $J$ high-temperature limit is reached above $T/J > 2$.

The charge susceptibility  is governed by a  single energy scale, {\it
i.e.} the  quasiparticle  gap  $\Delta_{qp}$.   This is  reflected  in
nearly perfect  scaling of $J\chi_c(T/J)$ for the  $N=8$ system in the
intermediate coupling  region  $1.2\le J/t\le  2.5$ (see  the inset of
Fig.~(\ref{chi}b)). Scaling is due to  the fact that the quasiparticle
gap  scales nearly linearly   with $J$, {\it i.e.}  $\Delta_{qp}\simeq
0.3J$ in this regime.  The  quasiparticle  gap remains finite even  at
small $J\sim 1.2$ where the spin gap disappears.  The scaling does not
persist up  to the strong coupling limit,  $J/t=10$ due to a crossover
regime where the  spin gap becomes  larger than the quasiparticle gap.
The  location of   the peak at   $T=T_{qp}$  seen  in the  temperature
dependence of  the charge susceptibility  curves matches  the value of
the quasiparticle gap  $T_{qp}=\Delta_{qp}$ obtained from measurements
of  the    doping    dependence of   the   chemical   potential   (see
table~(\ref{gaps})).  Despite a smaller  system size we  believe, that
near or at  zero doping, $N=8$  system shows less  finite-size effects
then the $N=10$    when calculating quasiparticle   properties  of the
system. The reason is that the $N=8$ noninteracting fermion system has
a  six-fold degenerate level  at zero energy  which  overlaps with the
value of  the chemical potential   at zero doping.  In  contrast,  the
$N=10$ noninteracting system has a large gap at $\mu$.  The scaling is
therefore  less obvious for  the $N=10$ system size, however locations
of  the peaks  nevertheless approximately   scale  with $J$  and  peak
positions $T_{qp}$ approximately match the quasiparticle gaps obtained
from the chemical potential curves for $J=1.6, 2.0,$ and $ 2.5$.

\subsubsection{Finite doping}

Spin and charge  susceptibilities,  presented in Fig.~(\ref{chidop}a),
at small doping $\delta=0.1$ show similar high-temperature behavior as
in  the zero doping   case.    At high-temperature,  $\chi_s$  follows
Curie-like $1/T$ behavior, predicted by the high-temperature expansion
given by Eq.~(\ref{chis}).  As in the zero-doping case, susceptibility
curves   calculated   for    different   $J/t$   show   scaling  above
$T/J>T_c/J\sim 0.6$.  With  decreasing temperature $\chi_s$  reaches a
peak  at $T=T_s$ where $T_s$  is close to its  $\delta=0$ value.  Even
though the spin gap disappears  at finite doping and zero temperature,
at finite temperature remains of the spin  gap can be observed even at
finite  doping.   Similar results  were   observed in  one-dimensional
calculations   \cite{tsune1}.    With  further   decreasing  of    the
temperature $\chi_s$  first decreases  and then sharply   increases at
even lower  temperature.  In this  region the susceptibility curve can
be fitted to  a simple form  $\chi_s = \delta  C /T$.  This Curie-like
form suggests that finite doping produces nearly free localized spins.
In contrast to one-dimensional   results \cite{tsune1}, we  found that
for intermediate values of the Kondo coupling $J/t=1.6, 2.0$ and $2.5$
the local  moment is reduced and  equals $C\sim 0.18$.   In the strong
coupling limit, $J/t=10$,  the local moment  reaches its maximum value
$C = 0.25$.  In the extreme  low-temperature limit spin susceptibility
should saturate either due to RKKY  interaction between localized spin
for small values of $J/t$ or due to Kondo screening effects for larger
$J/t$.   This  effect  can  be seen  in  the  case   of larger doping,
$\delta=0.2$, where   for  intermediate  Kondo coupling,  {\it   i.e.}
$J/t=1.6,2.0$  and   2.5  susceptibility curves show    less divergent
behavior as in the $\delta=0.1$ case.
\begin{figure}[tb]
\begin{center}
\epsfig{file=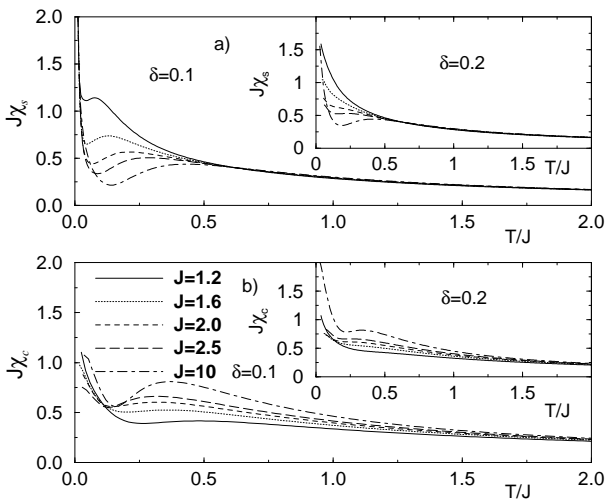,height=80mm,angle=-90}
\end{center}
\caption{ Spin  a) and charge   b) susceptibilities $J\chi_s$, $J\chi_c$
vs. $T/J$  at $\delta=0.1$ and 0.2.  Legends, given in b) apply also
for a) and both insets. }
\label{chidop}
\end{figure}
The    charge   susceptibility      at  $\delta=0.1$,  presented    in
Fig.~(\ref{chidop}b),   follows  $1/T$ behavior  at  high-temperature,
predicted by the high-temperature expansion given by Eq.~(\ref{chic}).
Susceptibility  reaches a local  maximum around  $T=T_{qp}$. For small
doping $T_{qp}$ overlaps with its value at $\delta=0$. At even smaller
temperature  we     observe a   sharp   increase   in   $\chi_c$.   At
zero-temperature charge   susceptibility should equal   the density of
states at the  Fermi  energy which is    further proportional to   the
quasiparticle mass.  The sharp increase of $\chi_c$ therefore suggests
that quasiparticles are massive.

\subsection{Specific heat}
Results for the specific heat  $c_v$ are shown in Fig.~(\ref{cv})  for
various values  of  $J/t$.  In  Fig.~(\ref{cv}a)  we show  $c_v$  as a
function of $T/t$. At $J/t=0$, $c_v$ has a peak at finite temperatures
which originates from the specific  heat of free conduction electrons.
The contribution of localized  noninteracting spins is nonanalytic and
proportional to   $T\delta(T)$ where $\delta(x)$  is a delta-function.
At small  values of Kondo coupling,  {\it e.g.   } $J/t=0.8, 1.2, 1.6$
and 2.0  we   observe  two    peaks  in  the    specific heat.     The
low-temperature peak that has emerged from the nonanalytic function at
$J/t=0$  is a contribution  of  the spin excitations.  With increasing
$J/t$ it   shifts  towards higher  temperatures    and broadens.   The
broadening is due to the spin excitation spectrum that has a bandwidth
of some effective  $J_{eff}$ which increases with  the strength of the
Kondo coupling.   The  broad peak,  that   originated  from the   free
electron  band at  $J/t=0$,   shifts towards higher temperatures  with
increasing  $J/t$   and becomes  even broader.   This   is due  to the
interplay of  band   effects and the  charge   gap that develops  with
increasing $J/t$.  The two  peaks above $J/t=2.5$  merge into a single
peak which in the strong coupling limit  scales linearly with $J$.  At
$J/t=10$  specific    heat   closely  follows   analytical  prediction
calculated  within the atomic  limit as seen in Fig.~(\ref{cv}b) where
$c_v$ is presented as  a  function of  $T/J$.    Our results for   the
two-dimensional lattice  are in many respects   similar to results for
the   one-dimensional   lattice   obtained     by  the   DMRG   method
\cite{shiba2,shiba}.
\begin{figure}[tb]
\begin{center}
\epsfig{file=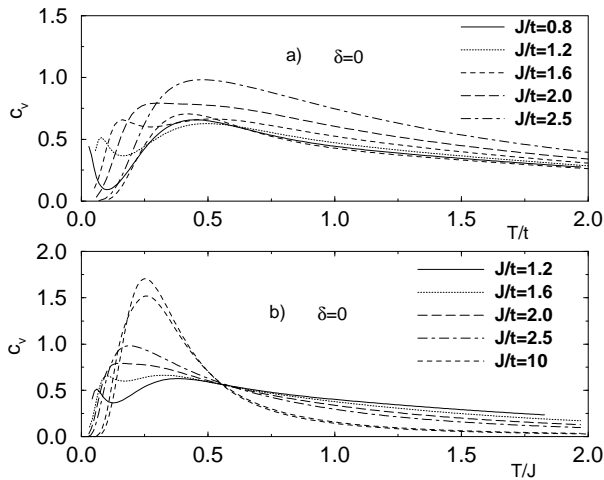,height=80mm,angle=-90}
\end{center}
\caption{  Specific heat at   $\delta=0$ $c_v$  a)  vs. $T/t$   and b)
vs. $T/J$ calculated  on a system of $N=10$.    The faint dashed  line
represents  the free  electron result a)   and result  obtained in the
atomic limit b).  }
\label{cv}
\end{figure}

At finite doping, $\delta=0.1$ shown in Fig.~(\ref{cvf}a), small peaks
due to the spin gap are  still visible at  small temperatures. At even
larger  doping,  $\delta=0.2$ shown  in  Fig.~(\ref{cvf}b), new  peaks
emerge   at  very  small  temperatures    (see also    the  inset  of
Fig.~(\ref{cvf}b)). This  peaks shift towards larger temperatures with
increasing Kondo coupling.   The shift is approximately  quadratic  in
$J$.  We  believe that the  same low  energy scale  that gives rise to
these peaks is responsible for saturation of spin susceptibility seen
in the intermediate coupling $J/t$ and $\delta=0.2$.
\begin{figure}[tb]
\begin{center}
\epsfig{file=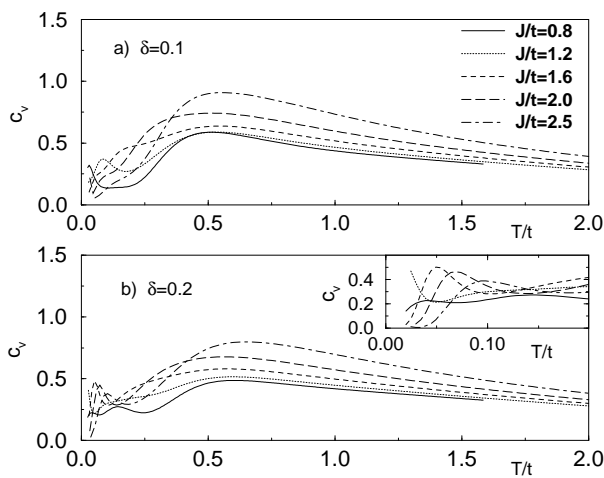,height=80mm,angle=-90}
\end{center}
\caption{ Specific heat $c_v$  $T/t$ at a) $\delta=0.1$ and b)
$\delta=0.2$ calculated
on a  system  of $N=10$. The inset  represents the expanded
low-temperature region in b). }
\label{cvf}
\end{figure}

\section{Conclusions}

We have  investigated   finite-temperature  properties  of   the Kondo
lattice model on small square lattices.  The chemical potential at low
temperatures shows nonanalytic behavior as a  function of doping.  The
jump in the  chemical potential is  a consequence of the quasiparticle
gap at  zero  doping.    The quasiparticle gap   scales  approximately
linearly  with the  Kondo    coupling, $\Delta_{qp}\sim 0.3J$ in   the
intermediate coupling regime, {\it i.e.} $1.6 \le J/t\le 2.5$. Similar
scaling  was  recently   found  by  quantum  Monte   Carlo simulations
\cite{assaad} which furthermore show  that scaling persists even below
the  transition to the  AFM state, {\it   i.e.} $J/t\sim 1.4$.  In the
strong   coupling regime the  quasiparticle  gap again scales linearly
with $J$ as $\Delta_{qp}\sim 0.75J$  which is not too surprising since
in this  case physics becomes local  and  $J$ is then the  only energy
scale in  the  system. The  crossover  between the two  linear regimes
occurs when the spin gap overcomes  the quasiparticle gap, {\it. i.e.}
at $J/t>2.5$. Interestingly, calculations   in one dimension show  two
distinct linear regimes for the charge gap vs.  $J/t$ \cite{shibata}.

Two temperature scales   govern  the temperature  dependence  of  spin
susceptibility in   the intermediate coupling  regime.    One scale is
given by $T_c/J\sim 0.6$ above which we find almost perfect scaling of
curves calculated for a wide  range of  coupling  strengths $J$.   One
possible explanation of this unusual scaling is that $T_c$ is governed
by  the charge gap $\Delta_c$.  We  have shown in the previous section
that  the quasiparticle gap at  intermediate  coupling scales linearly
with  $J$    as $\Delta_{qp}\sim  0.3J$.     Assuming $\Delta_c   =  2
\Delta_{qp}$ we get a linear scaling for the charge gap, $\Delta_c\sim
0.6J$, which agrees with the value for  $T_c$.  At lower temperatures,
$T\sim  T_s$,  physics at zero  doping  is  governed  by the  spin gap
$\Delta_s$.     This      remains      true     as         long     as
$\Delta_s<\Delta_{qp}$.  Near   the   strong coupling  regime the
opposite becomes true,   then the  quasiparticle  gap  determines  the
low-temperature physics of the spin susceptibility.

The charge susceptibility shows  substantially different behavior than
the spin susceptibility. On a smaller system of  $N=8$ we found almost
perfect scaling  with $J$ in  the  whole temperature  range within the
intermediate Kondo coupling region. This result suggests that a single
energy scale,  identified  as  a quasiparticle  gap   $\Delta_{qp}\sim
0.3J$, governs the physics of the charge response of the system.

At finite doping  we  found  a new  energy scale.  It is reflected
in the  saturation of  divergent $1/T$  behavior in  $\chi_s(T)$ at
small temperatures and in the appearance of low-temperature peaks in
the specific   heat $c_v$    at $\delta=0.2$.   Approximate  quadratic
scaling of the position of the peaks in $c_v$ with the Kondo coupling
strength suggests,  that this energy scale  could be attributed to the
RKKY interaction between uncompensated spins at finite doping. 

One of the authors, J.B. is grateful for the hospitality and financial
support of Los Alamos National Laboratory  where part of this work was
performed. We also acknowledge the support of Slovene-American grant by
the Slovene Ministry of science. 

\begin{table}[tb]
\caption{The   quasiparticle  gap $\Delta_{qp}/J$  and  peak positions
$T_{qp}/J$ and  $T_s/J$ both presented  in units of the Kondo coupling
$J$.  Quasiparticle  gaps were obtained  from the limit $\mu(\delta\to
0,T\to 0)$.  Spin susceptibilities  $\chi_s$ can be in the temperature
interval $T/J<0.6$ and  for  $J/t\leq 2.5$ well  fitted  to a simple  form
$\chi_s = C/T  \exp(-T_s/T) $ where the effective  moment $C  $ varies
from $0.25$ for small $J/t$ to $0.5$ at large $J/t$.  Estimated errors
where not otherwise specified are within 5\%.}
\begin{tabular}{|c|ccc|}
$J/t$ $(N=8)$&  $\Delta_{qp}/J $ &$ T_{qp}/J$ & $T_s/J$\\ \hline 
 1.2  & 0.27 & 0.31  & 0.08 $\pm$ 0.04\\
 1.6  & 0.28 & 0.30  & 0.13 $\pm$ 0.03\\
 2.0  & 0.30 & 0.31  & 0.18 $\pm$ 0.03\\
 2.5  & 0.32 & 0.31  & 0.31 $\pm$ 0.03\\
10.0  & 0.56 & 0.38  & 0.47 $\pm$ 0.04
\end{tabular}
\begin{tabular}{|c|ccc|}
$J/t$ $(N=10)$&   $\Delta_{qp}/J $&$T_{qp}/J$ & $T_s/J$\\ \hline 
 1.2   & 0.40 & 0.49  & 0.09 $\pm$ 0.04  \\
 1.6   & 0.30 & 0.35  & 0.15 $\pm$ 0.03  \\
 2.0   & 0.29 & 0.32  & 0.23 $\pm$ 0.03  \\
 2.5   & 0.31 & 0.32  & 0.30 $\pm$ 0.03  \\
10.0   & 0.56 & 0.38  & 0.47 $\pm$ 0.04 
\end{tabular}
\label{gaps}
\end{table}


\end{document}